# CHOOSING THE RIGHT RELATIVITY FOR QFT

**Leonardo Chiatti**

AUSL VT Medical Physics Laboratory
Via Enrico Fermi 15, 01100 Viterbo, Italy

> The reforms of our spacetime world, made necessary by the theory of relativity and characterized by the constant $c$, the uncertainty relations which can be symbolized by Planck's constant $h$, will need to be supplemented by new limitations in relation with the universal constants $e$, $\mu$, $M$. It is not yet possible to foresee which form these limitations will have.
>
> W. Heisenberg, 1929

**Summary**
When speaking of the unification of quantum mechanics and relativity, one normally refers to special relativity (SR) or to Einstein's general relativity (GR). The Dirac and Klein-Gordon wave equations are an example of unification of quantum concepts and concepts falling within the domain of SR. Current relativistic QFT derives from the transcription of these equations into second quantization formalism. Many researchers hope that the unification of QFT with GR can solve the problems peculiar to QFT (divergences) and to GR (singularities).
In this article, a different strategy is proposed, in an informal manner. Firstly, emphasis is laid on the importance of the transaction notion for quantum theories, introduced by Cramer in the eighties and generalized by the author of this article. In addition, the unification is discussed of QFT not with SR or with GR, but with their "projective" extensions (PSR, PGR respectively) introduced between 1954 and 1995 by Fantappié and Arcidiacono. The existence emerges of new fundamental constants of nature, whose significance is discussed. It is assumed that this new context can be a suitable background for the analysis of specific QFT (quark confinement, divergences, α-quantization of elementary particle masses and decay times) and PGR (gravitational collapse) problems.

## 1. INTRODUCTION

The unification of quantum mechanics (QM) with a given theory of relativity requires that covariant wave equations be formulated for the symmetry group of that particular theory. For example, the Schrödinger equation for a single particle with zero spin is covariant with respect to the Galilei group of Galilean relativity; the Dirac and Klein-Gordon equations are covariant with respect to the Poincaré group of Einsteinian special relativity (SR).
These equations are then transcribed in the second quantization formalism, and their physical meaning changes; now they are no longer equations of motion for an individual particle of a given type (electron, meson, etc.); rather, they become dynamic equations for the field of all particles of that given type (electronic, mesonic field, etc.). This leads to the quantum field theory (QFT) and we have Galilei-covariant QFTs, Poincaré-covariant QFTs, etc.

Normally, the Galilei-covariant QFT is called the "non-relativistic quantum field theory", while the Poincaré-covariant QFT is called the "relativistic quantum field theory". This conventional usage must not, of course, mislead. On the one hand, all QFTs are "relativistic", in the sense that they are compatible with a certain theory of relativity; on the other hand, the possible theories of relativity are certainly not confined to Galilean relativity and SR. SR (which contains the Galilean limit case $c \to \infty$, where $c$ is the limit speed) can indeed be generalized in two distinct ways.

Moving from Minkowski metrics to more general Riemann metrics, including gravitational field effects, Einstein's general relativity (GR) or a similar metric theory of gravitation is obtained.

On the other hand, one may wonder whether SR admits of an extension within the framework of a four-dimensional holonomous geometry. Fantappié showed in 1954 (1) that this extension exists and is unique, in the sense that additional, wider-ranging extensions are not possible if the holonomy and the four-dimensionality are to be maintained. The relativity thus obtained contains a limit speed ($c$) and a duration limit ($t_0$); its invariance group is the de Sitter group, whence it is called de Sitter relativity or Fantappié-Arcidiacono special projective relativity (PSR), from the names of the researchers who first developed it between 1954 and 1995 (2-19).

In the limit $t_0 \to \infty$ PSR collapses in the usual SR. Furthermore, it is possible to construct a projective general relativity (PGR) in which the effects of the gravitational field are described by means of an anolonomous geometry and which returns to being the usual GR in the limit $t_0 \to \infty$. The relationship existing between PGR and PSR is almost identical to that existing between GR and SR. PGR was developed by Arcidiacono starting from 1964 (17-25).

One can therefore pose oneself the problem of writing the QM wave equations in such a way that they are compatible with PSR rather than with SR. This problem has been addressed and solved, at least as far as the PSR generalization of the Klein-Gordon and Dirac equations is concerned, by Notte Cuello and Capelas de Oliveira (26-29). The subsequent passage to second quantization has been studied by a number of researchers, for example Chernikov and Tagirov (30), Bros *et al.* (31) for scalar fields; Takook (32) for free spinorial fields.

Once that SR-covariant or PSR-covariant equations are available, the passage to the corresponding GR-covariant or PGR-covariant equations requires that conventional wave operators be substituted by more general operators that depend on the metric tensor. For example, in the passage from the SR-covariant Klein-Gordon equation to that valid in GR for a curved space, the D'Alembertian becomes $g^{\mu\nu}\partial_\mu\partial_\nu$, where the symbols have their usual meaning.

Based on what has been said up to now, it would seem therefore that the problem of unifying QFT with SR and GR, or alternatively with PSR and PGR, has to all intents and purposes been solved, posing only difficulties, if any, of a mathematical nature.

As is also known to non specialists, the problem of unifying QFT with relativity is actually much more complex than that of identifying a covariant formalism. Difficulties peculiar to QFT (divergences) and to GR (singularities) exist which many researchers [e.g. see references (33-35)] hope can be solved by a more extensive unification which includes the specifically quantum aspects of gravitational field. In any case, opinions do not agree on how a quantum theory of gravitational field ought to be constructed. It is difficult to combine nonlocal aspects of QM and QFT with the fundamental nature of the spacetime representation assumed by relativistic theories.

Precise choices will be made in this article. First of all, the discussion will chiefly concern the unification of QFT with projective relativity. Furthermore, this unification will not be discussed at the formalism covariance level but, following a suggestion by Bohm (36), at the *ontological* level; in other words, the attempt will be made to identify *significant order structures* in treating problems concerning divergences and singularities. The identification of these structures will require, as we shall see, both a particular ontology of quantum processes and the action of a new fundamental constant of nature which is $t_0$. This constant is infinite in conventional SR, so that in relativistic QFT in the ordinary sense (SR, GR) the structures that we shall describe do not exist.

For what concern the ontology adopted here for quantum processes, it will be that based on the *transaction* concept, originally developed by Cramer (37-39) and referred to here in a more general form (40-42). We shall thus be led to the problem of the possible existence of a maximum and minimum duration of the transaction. In a PGR universe the maximum duration of a transaction coincides with the chronological distance of the generic pointevent from the geometrical horizon, which is $t_0$. A possible minimum duration can be identified, consistently with the available empirical material concerning the elementary particles, with the Caldirola[1] "chronon" $\rho/c$, where $\rho$ is the classical electron radius. An important empirical fact is that the relation between the two durations, independent of cosmic time, depends solely on the fine structure constant $\alpha$. In other words, the electromagnetic fine structure constant links the cosmos with elementary particles. In the realm of elementary particles, the mass $m/\alpha$ (where $m$ = the mass of the electron) plays the role of a quantum of mass, which can well be seen in statistical studies on mass and lifetime spectra of leptons and hadrons. These quantities are quantized according to rules which contain the electromagnetic fine structure constant $\alpha$.

On the other hand, $\rho$ becomes the radius of the hadronic horizon within which gluons and quarks are confined. It is shown that the spacetime of color interaction processes which have no links with the external world is a PSR microuniverse in which the chronon plays the role of $t_0$. Brief reference is made, based on these assumptions, to a geometrical model of the strong fine structure constant; it is in qualitative agreement with available experimental data.

The introduction of the chronon leads on the one hand to the possibility of a finite QED as early as in first quantization, as recently shown by Kawahara, and on the other to the possibility, merely a conjecture for now, of eliminating singularities in PGR.

## 2. TRANSACTIONS

We assume that the only truly existent "thing" in the physical world are the events of creation and destruction (or, if one prefers, physical manifestation and demanifestation) of certain qualities. In the language of QFT these events are the "interaction vertices", while the different sets of manifested/demanifested qualities in the same vertex are the "quanta".

As an example, in a certain vertex a photon (E, **p**, **s**) can be created, where E is the (created) energy of the photon, **p** is the (created) impulse of the photon and **s** is the (created) spin of the photon. In a subsequent event this photon can be absorbed and this corresponds to a packet of properties (-E, -**p**, -**s**) where –E is the (absorbed) energy of the photon, - **p** is the (absorbed) impulse of the photon and – **s** the (absorbed) spin of the photon; (-E, -**p**, -**s**) is the absorbed, i.e. destroyed, photon. It is assumed that the first event chronologically precedes the second, and that E is positive. A process is therefore being described in which a positive quantity of energy is created and then destroyed. In an absolutely symmetrical manner, it can be said that we are describing a process which proceeds backwards in time and during which a share of negative energy –E is firstly created (in the second event) and then destroyed (in the first event). The two descriptions are absolutely equivalent.

There are two reasons for which, in the process described, one cannot have E < 0, i.e. the propagation of a positive energy photon towards the past. A photon with E > 0 which retropropagates towards the past, yielding energy to the atoms of the medium through which it travels (such as, say, a photon X which ionizes the matter through which it travels), is seen by an observer proceeding forward in time as a photon with E < 0 which *absorbs* energy from the medium (43). This photon would be spontaneously "created" by subtracting energy from the medium, through a spontaneous coordination of uncorrelated atomic movements which is statistically implausible, and has never been observed experimentally.

From a theoretical point of view, the probability of the occurrence of a creation/destruction event for a quantum $Q$ in a pointevent $x$ is linked to the probability amplitude $\Psi_Q(x)$, which can be a spinor of any degree. Each component $\Psi_{Q,i}(x)$ of this spinor satisfies the Klein-Gordon quantum relativistic equation $(-\hbar^2 \partial^\mu \partial_\mu + m^2 c^2) \Psi_{Q,i}(x) = 0$, where $m$ is the mass of the quantum[2]. At the non-relativistic limit, this equation becomes a pair of Schrödinger equations (44):

$$-\frac{\hbar^2}{2m} \Delta \Psi_{Q,i}(x) = i\hbar \, \partial_t \, \Psi_{Q,i}(x) \qquad (2.1)$$

$$-\frac{\hbar^2}{2m} \Delta \Psi^*_{Q,i}(x) = -i\hbar \, \partial_t \, \Psi^*_{Q,i}(x) \qquad (2.2)$$

---
[1] We omit here the 2/3 factor which appears in Caldirola's original definition.

The first equation has only retarded solutions, which classically correspond to a material point with impulse **p** and kinetic energy E = **p·p**/2m > 0. The second equation has only advanced solutions, which correspond to a material point with kinetic energy E = - **p·p**/2m < 0. Thus there are no true causal propagations from the future.

One may wonder whether eq. (2.2) can lead to hidden advanced effects. The answer to this question is affirmative. To understand this topic, one ought to reconsider the photon example seen above. The creation of the E > 0 energy followed by its subsequent absorption and, conversely, the creation of a –E < 0 energy preceded by its destruction are clearly two different descriptions of the same process. This however is true so long as the interaction events are considered, i.e. the true substance of the physical world.

From the point of view of the dynamic laws for the probability amplitudes of these events, matters are quite different, however. The creation of quality $Q$ is associated with the initial condition for $\Psi_{Q,i}(x)$ in eq. (2.1); the destruction of quality $Q$ is associated with the "initial", actually the final, condition for $\Psi'_{Q,i}(x)$ in eq. (2.2). The prime is justified by the fact that the two conditions are generally different and therefore generate different solutions for the two equations, which are not necessarily mutual complex conjugates.

It is a fundamental fact that the destruction event is not described by eq. (2.1) and that the creation event is not described by eq. (2.2); this remains true even if the Hamiltonians of interaction with the remaining matter are introduced into the two equations. Thus, at the dynamic laws level, the process of the creation and destruction of $Q$ is completely described solely by the ring:

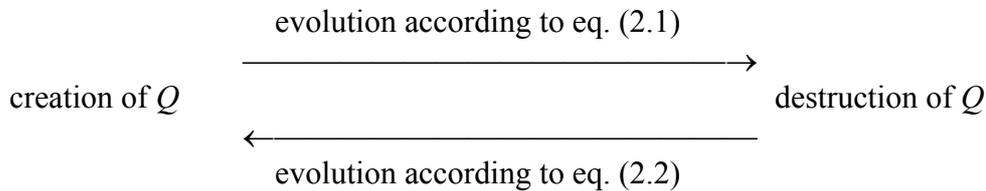

and not only by the upper or lower half-ring.

---

[2] It is necessarily positive, because it is a measure of the energy which must be released in order to create the quantum; a physical entity can certainly not be created in the vacuum by subtracting energy from it. For a more precise justification of the topics set forth, please see ref. (42).

More generally speaking, we shall have at $t = t_1$ the event of the creation-destruction of a quality $Q$ ( $|Q><Q|$ ) and at $t = t_2$ the event of the creation-destruction of a quality $R$ ( $|R><R|$ ). These two processes will be linked by a time evolution operator $S$ according to the ring:

$$
\begin{array}{ccc}
|Q> & <Q| & t = t_1 \\
S \downarrow & \uparrow S^+ & \qquad (2.3) \\
|R> & <R| & t = t_2
\end{array}
$$

In other words, $|Q>$ is transported from $S$ into $|Q'>$ and projected onto $<R|$, $|R>$ is transported by $S^+$ into $|R'>$ and projected onto $<Q|$. The amplitudes product :

$$<R|S|Q><Q|S^+|R> \;=\; |<R|S|Q>|^2$$

is immediately obtained, which is the probability of the entire process. If quality $Q$ is constituted by a complete set of constants of motion then $R = Q$ and this is the type of process which can describe the propagation of a photon-type quantum, otherwise it is the generic process of the creation of a quality $Q$ causally linked (by means of $S$) to the destruction of a quality $R$. Moving to the representation of the coordinates, by substituting bras and kets with wavefunctions, we once again obtain as a particular case the result already seen with the non-relativistic expressions (2.1), (2.2).
For the ring process described here, Cramer (37-39, 45) coined the term *transaction*[3]. The transaction exists if the propagation $S^+$ is as "real" as the $S$; to ascertain this, one must see whether experimental situations exist in which the initial condition $|Q>$ and the final one $<R|$ are connected in a nonlocal way. As is well known, the answer is affirmative: the phenomena EPR and GHZ (46) are highly valid examples of quantum-mechanical predictions which violate locality. In the case of EPR phenomena there is now confirmed solid experimental evidence (47-50).

Some remarks.

1) If one just considers only the propagation $S$, the quantum nonlocality vanishes and one therefore cannot explain its origin without introducing additional hypotheses to the formalism in use.

2) In the same hypothesis, one cannot explain the destruction of the "quantum state" as a phenomenon that takes place at a defined instant $t_2$. Well known paradoxes, such as that of Schrödinger's cat (51), derive from this.

3) From an algebraic point of view, the transactional ring is a sort of identity operator, because $SS^+ = S^+S = 1$ and the qualities $Q$, $R$ are simultaneously created and destroyed. One has the impression that every quantum process (therefore all matter) and time itself are emitted from an invariant substratum and re-absorbed within it. This substratum cannot be observed because it is invariant and outside of space and time, which originate from it. A sort of "motionless motor".
   The transactional ring may be described by circular inference rules which establish a self-consistency rather than, as in traditional logic, by linear inference rules which establish a deduction. A circular logic, not a linear one, is applied to a ring; it appears therefore that individual quantum processes are self-generated in a non-causal manner through an extra-spacetime mechanism.

---

[3] Cramer introduced this concept in the framework of classical Wheeler-Feynman electrodynamics. The author here proposes to redefine it in a purely quantum context, in accordance with references (40-42).

Self-generation implies acausality. At the same time, however, the creation/destruction events that occur at $t = t_1$, $t = t_2$ take place because of interactions with other rings. This implies the existence of rules on how the rings connect and therefore the acausality does not turn into complete arbitrariness. This appears to be a natural and entirely convincing explanation of the simultaneous presence of causality and acausality in quantum processes.

4) The energy is propagated only in one time direction, and the causal effects thus proceed from the past towards the future.

To sum up, the two extreme events of a transaction correspond to two reductions of the two state vectors which describe the evolution of the quantum process in the two directions of time. They constitute the "**R** processes" (**R** stands for *reduction*) of the Penrose terminology (33-35) and, from our perspective, are the only real physical processes. They are constituted of interaction vertices in which real elementary particles are created or destroyed; these interactions are not necessarily acts of preparation or detection of a quantum state in a measurement process.

The evolution of probability amplitudes in the two directions of time constitutes, in Penrose terminology, a **U** process (where **U** stands for *unitary*). From our ontological viewpoint, **U** processes are not real processes: both the amplitudes and the time evolution operators which act on them are mathematical inventions whose sole purpose is to describe the causal connection between the extreme events of the transaction, i.e. between **R** processes. This connection is possible because the two events derive from the transformation of the same aspatial and atemporal "substratum". As a specific consequence of this assumption, *all virtual processes contained in the expansion of the time evolution operator are deprived of physical reality*.

According to this approach, therefore, the history of the Universe, considered at the basic level, is given neither by the application of forward causal laws at initial conditions nor by the application of backward causal laws at final conditions. Instead, it is assigned as a whole as a complete network of past, present and future **R** processes. Causal laws are only rules of coherence which must be verified by the network and are *per se* indifferent to the direction of time.

As an example of a transactional network, let us consider a process well known in QFT, constituted by the decay of a microsystem, prepared in the initial state 1, into two microsystems 2, 3 which are subsequently detected. The preparation consists of the destruction of quality 1 [which we shall indicate with ( 1 |] which closes the transaction which precedes it, and of the creation of quality 1 [which we shall indicate with | 1 )] which opens a new transaction. It will be represented by the form | 1 )( 1 |.

The decay consists of the destructions of qualities 2, 3 which close the transaction that began with the preparation, and of the creations of qualities 2, 3 which open a new transaction which will be closed with the detection of microsystems 2, 3. It will be represented by the form [| 2 )| 3 )] [( 2 |( 3 |].

The detection of microsystems 2, 3 will be constituted by the destructions of qualities 2, 3 which close the transaction that began with the decay, and by the creations of qualities 2, 3 which open subsequent transactions. It will be represented by the interaction events | 2 )( 2 | , | 3 )( 3 |.

The double transaction described here corresponds to the process usually associated with the probability amplitude $< 2, 3 \mid S \mid 1 >$.

Another example is Young's classical double slit experiment. The preparation of the initial state of the particle can be represented by the form | 1 )( 1 |, following the same reasoning as in the previous case. Instead of the decay, here we have the crossing of slits 2 and 3, i.e. the interaction between a particle and a double slit screen represented by [| 2 )| 3 )] [( 2 |( 3 |]. Instead of the detection of the two particles created in the decay, here we have the sole event of the detection of the particle on the second screen at a certain position 4, i.e. : | 4 )( 4 |. Two transactions are involved: the first starts

with the preparation of the particle and ends with its interaction with the first screen; the second begins with this second interaction and ends with the interaction of the particle with the second screen. The latter interaction then constitutes the beginning of the following transaction. The process is that which corresponds to the probability amplitude $< 4 \mid S \mid 1 >$.

We note that the forward time evolution of the amplitudes, represented with $S \mid 1 >$, contains both the kets $\mid 2 >$, $\mid 3 >$; nevertheless, processes relating to the passage through the individual slit $a$ (where $a = 2, 3$) do not exist. Such processes would require an intermediate event represented by $\mid a )( a \mid$, which effectively does not take place. It is in this sense that processes that can be associated with compound probability amplitudes $< 4 \mid S \mid a > < a \mid S \mid 1 >$ are "virtual" and not real. The process of the crossing of one of the two slits becomes real when the other slit is closed.

### 3. TRANSACTIONS ON A PGR CHRONOTOPE

Let us consider, in PGR spacetime, an observer placed at the chronological distances $t'$, $t''$ from the two singularities of the big bang and possible big crunch, respectively. These singularities - if very far from the observer - appear to him to be "crushed" against the two surfaces, past and future, of the de Sitter horizon (52,53). The chronological distance of these two surfaces from a generic observer is $t_0$, so that $t'$, $t'' < t_0$. If there is no big crunch, then $t'' = t_0$.

Therefore, this observer finds that the interval between the extreme events (creation and destruction) of any transaction, i.e. its total duration, cannot exceed $2t_0$. Thus, there is a maximum duration of transactions in PGR when these are observed on the generic observer's individual chronotope. We must bear in mind that in the limit constituted by PSR, this chronotope becomes the Castelnuovo chronotope (17-19).

Clearly, there is no accurate observational value available for $t_0$. From the relation $t_0 = 1/H_0$, valid in PSR, between this constant and the current value $H_0$ of Hubble's constant, it can be estimated that $r = ct_0 \approx 10^{28}$ cm.

One instinctively wonders whether a minimum duration exists for transactions. If this duration exists, all physical phenomena of shorter duration than it must link events which are not extremes of transactions and which therefore **are not R processes**. These phenomena must therefore, according to Penrose's terminology, constitute aspects of a **U** process. In terms of quantum physics everyday language, the linked events are therefore extreme vertices of "virtual" propagations that exist solely as terms of the expansion of the time evolution operator of the system being studied. Such propagations would be relative to quanta that cannot be detected by an observer at infinity, as they are limited to a distance equal to $c$ times ($c$ = speed of light in vacuum) the minimum duration in question.

The question can therefore be reformulated thus: do quanta exist whose propagation is limited below a certain spatial distance and which have never been observed as free? It is a well known fact that the experimental answer is affirmative: these are quarks and gluons, and their propagation is limited to a distance roughly equal to $e^2/mc^2 \approx 10^{-13}$ cm, the classical radius of an electron, which is also the maximum hadronic radius ($e$ = elementary charge, $m$ = electron rest mass). The time interval in question is, therefore, roughly $\theta_0 = e^2/mc^3 \approx 10^{-23}$ s and coincides - unless a 2/3 factor - with the Caldirola "chronon" (54).

The relation between $\theta_0$ and $t_0$ is the inverse of Dirac's number $10^{41}$, and it must be observed that in PGR $t_0$ is a constant, so that this relation does not depend on cosmic time. It can also be written in the alternative form suggested by Sternglass (55-57):

$$t_0 \approx 2^{(1/\alpha)} \theta_0 \ , \qquad\qquad\qquad (3.1)$$

where $\alpha$ is the electromagnetic fine structure constant. This relation cannot be formulated within the context of usual relativistic cosmology, because in it $t_0$ is infinite. On the other hand, by substituting $t_0$ with the current value of the age of the Universe, a relation is obtained that is dependent on cosmic time which, as is well known, is incompatible with experimental data (58).

One can of course object that the "numerology" represented by eq. (3.1) is merely a non significant coincidence. It must be noted, however, that equation (3.1) is not an isolated relation, as it is a special case of a much wider empirical law. Ramanna and Sreekantan (59) find the following duality formula between the mass $M$ and the half-life $\Lambda$ of an elementary particle:

$$\hbar/(Mc^2 \Lambda) = n/2^n \quad . \tag{3.2}$$

In this relation $n$ is normally very close to an integer number. The same type of relation is valid for β-emitting radionuclides, if $M$ is the mass of the neutron, or for α-emitting ones if $Mc^2$ is the bond energy of the emitted nucleus (59-65). For some interesting coincidences exhibited by $n$ index in the domain of elementary particles the reader is referred to Appendix.

We observe that in the case of an electron on a PGR chronotope, $M = m$ and $\Lambda = t_0$, as it is a stable particle with a duration equal to the maximum duration of the Universe. With this position, equation (3.2) is transformed into equation (3.1) assuming that $n = 137 \approx 1/\alpha$. One must remember that the electron is the lightest stable particle, if one excepts gauge quanta. The proton is stable but is heavier than the electron, while neutrinos are lighter but are created in oscillating superpositions of mass eigenstates and therefore, in this sense, are not stable. It is plausible, therefore, that the mass of the electron is a sort of "sample mass" on which the masses of the other particles are gauged in some way. We shall return to this issue below.

Equation (3.1) can give rise to doubts of the following sort. The value of $\theta_0$ is a property of elementary particles and, therefore, of the individual **R** processes which are substantially events. How can an individual event to be informed of a global property of the Universe such as $t_0$? The relation (3.1) must therefore be fortuitous.

However, one must remember that **R** processes are processes of creation-destruction of properties (including the internal quantum numbers of particles) which connect the physical world with an aspatial and atemporal substratum. At the level of this substratum, all the past, present and future **R** processes are therefore connected. A *cosmological* nonlocality is therefore present, in addition to the already acknowledged quantum nonlocality reflected in the phenomenon of amplitude entanglement, and equation (3.1) reflects this very cosmological nonlocality. This interpretation is suggested by Kafatos (66-68).

## 4. MICROUNIVERSES

Equation (3.1) suggests that the couple of fundamental constants $t_0$, $\theta_0$ (or, if one prefers, their ratio) defines a bridge between microphysics and cosmology. To understand better the nature of this "bridge" a deeper analysis is required of the role of the constant $\theta_0$. Every physical process shorter than $\theta_0$ must be a **U** process; on the other hand, a **U** process can be represented as a succession of a greater or smaller number of consecutive **U** processes with duration $\theta_0$. If A, B are two virtual events connected by a virtual process shorter than $\theta_0$ and $(x, y, z, ict)$ is their relative distance vector [projected on the chronotope tangent to the PGR manifold in one of them - say, A -] then one must have:

$$c^2 t^2 - x^2 - y^2 - z^2 \leq \theta_0^2 \tag{4.1}$$

and this condition must remain valid regardless of the frame of reference chosen in the neighbourhood of A. If equation (4.1) could be violated with an appropriate choice of frame, then in that frame A and B could be the extremes of a transaction, i.e. **R** processes. But since the nature of an **R** process is completely independent of the choice of reference, this result is impossible. In all inertial frames of reference physically compatible with the existence of the fundamental interval $\theta_0$ equation (4.1) must therefore apply. These references will be linked by trasformations of co-ordinates which leave equation (4.1) unchanged and these transformations form, as is known, the de Sitter-Fantappié group. The spacetime associated with this group is the de Sitter *microuniverse* having radius $\rho = c\theta_0$; its projection onto ordinary "external" spacetime constitutes a Castelnuovo microuniverse.

Thus, following this line of reasoning, the time evolution of the probability amplitude of a hadron, both in first (QM) and in second (QFT) quantization will be described by a sum of paths (each equivalent to a different "history" of the particle), each of which will be a succession of universe tubes having a time length $2\theta_0$ and spatial diameter $2\rho$. Each of these "elementary tubes" will be a de Sitter-Castelnuovo microuniverse.

In the case of leptons, these "elementary tubes" will still have a time length $2\theta_0$ while their spatial diameter will be zero. Leptons, indeed, are not composed of quarks that interact by exchanging gluons and therefore there are no spatially extended exchange processes inside leptons.

In these "particle microuniverses", events are internal to particles (or at the vertex of interaction between particles) and are not connected to observers located at spatial and temporal infinity. Fantappié-Arcidiacono PSR kinematics apply to them. Relation (3.1) connects macrocosm dimensions with those of the microuniverses. Alternatively, it can be said that it defines a transformation of scale from the PSR that is approximately valid in the macrocosm (PGR considered in the limit case of low densities of matter) to a PSR that is exactly valid in the microcosm.

The existence of a "scaled" de Sitter relativity for elementary particles implies various interesting consequences, which we shall examine briefly.

**4.1 Leptonic microuniverses**

The simplest case is undoubtedly that of leptons, because for leptons the only difference from the standard theory is the decomposition of each path, in the relevant path integral, into consecutive segments having duration $\theta_0$. In the context of first quantization, Kawahara (69) has recently proposed a modification of the Dirac equation that is compatible with this idea, while to our knowledge nothing has yet been done within the framework of second quantization formalism.

Kawahara introduces an evolution parameter *s* which satisfies the Breit-Wigner equation $dx^\mu/ds = \gamma^\mu$. Thus, if $\Psi$ is the electron wavefunction :

$$\frac{d}{ds}\Psi = \gamma^\mu \partial_\mu \Psi \quad . \tag{4.2}$$

As is well known, *s* locally coincides with proper time, given by the square root of the spacetime separation [ $g_{\mu\nu} = diag(+,-,-,-)$ ] :

$$s^2 = g_{\mu\nu} x^\mu x^\nu \quad , \tag{4.3}$$

though this does not hold true globally. Kawahara also assumes the existence of a particular interval $\delta s$ of *s* such that:

$$\overline{\Psi}(s+\delta s)\Psi(s+\delta s) = \overline{\Psi}(s)\Psi(s) \quad . \tag{4.4}$$

Starting from these premises, he obtains the following generalization of the Dirac equation (in natural units):

$$\left(i\gamma^\mu \partial_\mu - e\gamma^\mu A_\mu + V(\delta s) - \frac{2\pi n}{\delta s}\right)\Phi = 0 \quad , \tag{4.5}$$

where $\Psi = \Psi(x_\mu(s), s) = \Phi(x_\mu)\exp(iMs)$; $n = 0, \pm 1, \pm 2, \ldots$ and $V(\delta s) = V_1 + V_2 + \ldots$ with:

$$V_1 = +\frac{1}{2}\delta s\, e\left[\gamma^\mu \partial_\mu, \gamma^\nu A_\nu\right] \tag{4.5a}$$

$$V_2 = -\frac{1}{12}(\delta s)^2 e\left[\left[\gamma^\mu \partial_\mu, \gamma^\nu A_\nu\right], \gamma^\xi\left(i\partial_\xi + eA_\xi\right)\right] \quad . \tag{4.5b}$$

The physical reason of the last term of equation (4.5) is very clear: because of equation (4.2), the wavefunction must be periodical in $s$ with a period $M$ (where $M$ is the physical mass), if one is to have an isolated particle ($A_\mu = 0$) having a defined mass $M$. Only in this way can a common factor that is periodical in $s$ be eliminated from the two members of equation (4.2), thus obtaining a wavefunction $\Phi$ that is no longer explicitly dependent on this parameter. It follows from this that in the last term of equation (4.5) it must be $n = 1$ and $\delta s = 2\pi/M$. **After having performed this substitution**, the ordinary Dirac equation is once again obtained in the limit $\delta s \to 0$.

The last two terms of equation (4.5) constitute the renormalized mass. As one can see, in the limit $\delta s \to 0$ one has $V(\delta s) \to 0$, while the last term diverges. Now, if a "duration quantum" $\theta$ of the "electron" process exists, such divergence should actually appear for $\delta s \to k\theta$, where $k$ is of the order of unity. By substituting $\delta s$ with $\delta s - k\theta$ in the last term of equation (4.5), this becomes:

$$2\pi/(\delta s - k\theta) = 2\pi/[\delta s(1 - k\theta/\delta s)] \approx (2\pi/\delta s)[(1 + k\theta/\delta s)] = (2\pi/\delta s) + \alpha k (2\pi/\delta s),$$

where $\alpha = \theta/\delta s$. It follows that a percentage contribution equal to $ak$ must be added to the mass $(2\pi/\delta s)$ evaluated for $\theta = 0$, in order to take into account the finite value of $\theta$. Experimentally, $\alpha$ is an universal constant (electromagnetic fine structure constant). If $k = 1/\pi$, the additional term $\alpha kM$ at the mass evaluated for $\theta = 0$ is precisely the correction $\delta M$ which, after insertion in the contribution of $V_1$ to the magnetic moment [equation 53 of Karawaha's paper], provides the correct anomalous gyromagnetic ratio of the particle (evaluated at the first order). This is an alternative to Kawahara's reasoning, which instead obtains the same term by evaluating the electrostatic self-interaction with a procedure which is physically less clear.

Subsequently, Kawahara calculates the contribution to the anomalous magnetic moment from the term $V_2$ (higher-order radiative corrections). The entire procedure is free from divergences and renormalization is not necessary. Obviously, in the case of the electron $\theta$ assumes its highest value, which is $\theta_0$.

Kawahara does not go more deeply into the fact that from equation (4.2) and from the Breit-Wigner equation it follows automatically that $\Psi = \Psi(x_\mu(s), s)$; from this relation the dissimulated existence of virtual "paths" $x_\mu = x_\mu(s)$ emerges; for a closer analysis of this aspect, see Chiatti (42, 70). We only point out here that if a quantum of time $\theta$ exists and the classical limit is considered in which only an "average" path (the actual path) survives, then:

$$x_i[n\tau_0] - x_i[(n-1)\tau_0] = (\tau_0/2)\{u_i[n\tau_0] - u_i[(n-1)\tau_0]\} = (\tau_0/2)\Delta u_i[n\tau_0] \tag{4.6}$$

a relation in which the three components ($i = 1, 2, 3$) of position $x$ and of speed $u$ of the particle appear and $\theta = \tau_0/2$. Caldirola proposed, starting from the 1950s, a classical theory of the electron alternative to that of Dirac, coupling equation (4.6) with the following dynamic equation at finite differences:

$$(m/\tau_0)[\Delta u_i(\tau) + u_i(\tau)u_k(\tau)\Delta u_k(\tau)] = (e/c)F_{ik}(\tau)u_k(\tau) \ . \qquad (4.7)$$

He solved the system (4.6) + (4.7) for simple special cases [free motion, motion under a constant field, hyperbolic motion], always obtaining finite and consistent results (71-77). The Caldirola theory does not present "run away" solutions nor does it have problems with causality; it is a known fact that these difficulties are present, instead, in the approach based on the classical Dirac equation. The Caldirola $\theta$ quantum (Caldirola chronon) differs from that adopted here by a factor of 2/3.
Caldirola was the first to consider the electron as a de Sitter-Castelnuovo microuniverse (77), with an expansion/collapse cycle having a total duration of $2\theta_0$. In any case, the electron does not have a spatial extension but only a temporal one; this cycle, therefore, despite Caldirola's contention, has no "internal" dynamic effect on a free electron. In the following section, the case of the hadronic microuniverse is examined, whose radius coincides with the field of action of the colour force and whose spatial coordinates are those which express the relative positions of quarks (hadrons are spatially extended). In this microuniverse, dynamic effects connected with global space curvature are present.

**4.2 Hadronic microuniverses**
The idea of interpreting hadrons as spatially extended structures within a GR microuniverse having a radius $\approx c\theta_0$ was originally explored by Recami and his collaborators (78-89). In any case, we shall assume here that the hadronic microuniverse is a de Sitter-Castelnuovo PSR chronotope. The generic observer inside the Castelnuovo microuniverse would see the two surfaces of the de Sitter horizon, the past and future ones, at the respective chronological distances $-\theta_0$ e $+\theta_0$.
This Castelnuovo microuniverse can be considered as the geodetic projection of a de Sitter microuniverse on the ordinary spacetime [this was Caldirola's construction for electrons, entirely similar to Arcidiacono's for the PSR Universe]; in this second microuniverse, the temporal axis $t'$ is unlimited. If the tangential point of the projection coincides with the observer and $t$, $t'$ are the chronological distances from the observer in the two microuniverses, the Fantappié relation holds (17-19):

$$t' = \frac{\theta_0}{2} \log\left(\frac{\theta_0 + t}{\theta_0 - t}\right) \ . \qquad (4.8)$$

Let us then imagine that gluons motion is not free in the de Sitter microuniverse, but that they can be emitted by a quark only if they are subsequently absorbed by another quark. This is clearly a global, not a local constraint, which means that the interaction force between quarks does not depend on their distance $\xi$ on the de Sitter microuniverse. Since gluons move at the speed of light, $\xi = ct'$, where $t'$ is the time interval, counted on the de Sitter microuniverse, between the emission and the absorption of a gluon.
As a result of equation (4.8), the interquarkic force will therefore depend on the distance $d = ct$ travelled by a gluon in the Castelnuovo microuniverse. If we consider, in this microuniverse, a quark-observer placed at the origin, the gluons absorbed by it will be emitted by other quarks at an ever-accelerating rate, the closer they are to the surfaces of the de Sitter horizon; consequently, the force exerted on the quark-observer by other quarks will also grow indefinitely. In other words, the strong fine structure constant will depend on $d$ according to the obvious relation:

$$\alpha_{strong}(d) = \alpha_{strong}(0) \,|\, dt'/dt \,| = \alpha_{strong}(0) / [\,1 - (t/\theta_0)^2\,] =$$

$$= \alpha_{strong}(0) / [\,1 - (d/\rho)^2\,]. \tag{4.9}$$

It is now necessary to express $\alpha_{strong}$ as a function of the interaction energy $E_{int}$ exchanged in a hadronic strong interaction process. At first sight, one might be tempted to substitute, in equation (4.9), the uncertainty principle in the form $t = \hbar/E$ after having assumed that $\hbar/\theta_0 = E_0 = 70$ MeV. Thus, $(t/\theta_0)$ would be substituted by $(E_0/E) = \alpha_{strong}(E_{int})(E_0/E_{int})$ and one would have a third degree equation for $\alpha_{strong}(E_{int})$.

However, the uncertainty principle applies to the duration $T$ of the quantum state, not to the time interval $t$ necessary for an individual gluon to travel the interquark distance $d$. $T$ is the average time in which two quarks exchange a gluon and generally $T > t$. The $T \approx t$ case basically concerns the collision between two particles coming from infinity which immediately move away to infinity. In the quark case, originating from infinity and moving away to infinity are not possible, at least within a single stationary hadron.

The function $\alpha_{strong}(E_{int})$ can be constructed, using an analogy with relation (3.1). The duration $T$ of the process is the statistical mean of the length of the portion of timeline accessible to that process. One can therefore define the strong coupling constant by means of the relation:

$$T \approx 2^{(-k/\alpha_{strong})}\,\theta_0\,.$$

Thus, let $T/\theta_0 = (\hbar/\theta_0)(T/\hbar) = E_0/E_{int}$ ; we have, in the interval $E_{int} \geq E_0$:

$$\alpha_{strong}(E_{int}) \approx k/\log_2(E_{int}/E_0)\,.$$

Or, which is easier:

$$\alpha_{strong}(E_{int}) = 0{,}301\, k/\log_{10}(E_{int}/E_0)\,. \tag{4.10}$$

For $\alpha_{strong} = \infty$ we have $E_{int} = E_0$ from equation (4.10), and $d = \rho$ from equation (4.9). $\alpha_{strong}$ assumes a minimum value of 0 in equation (4.10), for $Eint = \infty$; it assumes, instead, a finite minimum value in equation (4.9), equal to $\alpha_{strong}(0)$, for $d = 0$. It is not possible to estimate $k$ or $\alpha_{strong}(0)$ based on the considerations developed here. On the basis of other considerations [(42), chapter XI] it can be deduced that $k = 1.134$. In Fig. 1 the comparison is shown between equation (4.10) and PDG2002 data. The theoretical curve has been plotted assuming the deduced value 1.134 for $k$; as can be seen, the agreement can be improved by applying a corrective factor - which is entirely empirical - of 1.16.

The difference between theoretical and experimental curves, which is greater at lower energy, is not surprising. Firstly, one must consider that the value of $\alpha_{strong}$ derived from equation (4.10) is a *bare* value and not an effective one; as such, it does not take into account the effects connected with the actual dynamics of virtual interaction. The experimental value, on the other hand, is obtained by elaborating experimental data through chromodynamic relations which presuppose effective coupling. We remark that $1.16 \approx 1 + 1/2\pi$ which remind us to the first term of some kind of perturbative expansion of $\alpha_{strong}$.

Another fundamental difference is that the colour forces are two-body forces in QCD, but it is not to be assumed that they are such in an approach such as this. If, as we have supposed, gluons are constrained to moving among quarks and "free" gluons do not exist, this means that a hadron is stationary only if the number of gluons circulating among its various quarks is stable. In other words, each quark must on average emit gluons at the same rate at which it absorbs them; thus its colour charge is greater the greater the distance from the other quarks. Let us consider, therefore, the interaction between a quark A, bonded with a quark B, and a quark C. If the interaction energy is high, the three quarks are very close; therefore the rate of emission of A towards C is not strongly affected by B. But at lower energy, if A is close to C yet far from B, the rate of emission of A towards C is dominated by the presence of B. Thus the AC coupling is greater than it would be if there were no B. This strengthening effect is not taken into consideration in equation (4.10); it is therefore obvious that the coupling constant derived from equation (4.10) is less than that estimated *globally*.

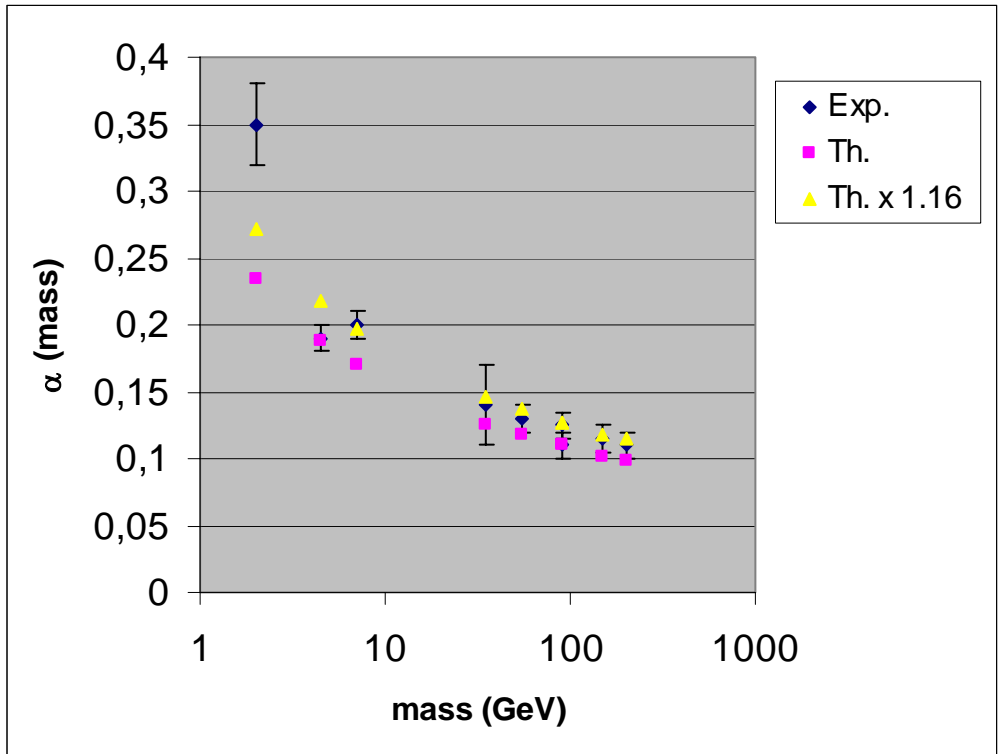

Fig.1 Strong coupling constant as a function of $E_{int}/c^2$. Experimental data from ref. (90), Fig. 9.2.

Let us remember that if $m$ is the mass of the electron, $E_0$ can be defined as $\alpha^{-1}mc^2$. A series of important connections thus begins to take shape: in the special case of the electron, the general relation (3.2) gives rise to relation (3.1); this defines $\theta_0$ starting from $\alpha$ and $t_0$, after which $m$ is defined by the relation $\hbar/\theta_0 = E_0 = \alpha^{-1}mc^2$. The twin relation (4.10), valid within the hadronic microuniverse, defines, instead, some important properties of strong interaction.

The constant $E_0 = 70$ MeV plays a very important role in the quantization of the mass spectrum of elementary particles. As early as 1952, Nambu had already remarked (91) a series of interesting relations such as $M_\pi = 2m/\alpha = 2E_0$ for the mass of the pionic triplet or $M_\mu = (3/2)m/\alpha = (3/2)E_0$ for the mass of the muon. We know today that for almost all elementary particles the mass $M$ is the $P$-th multiple of a quantum of mass $u$ very close to $E_0/2$. The integer number $P$ is odd for baryons and leptons, even for mesons and can be correlated with the internal quantum numbers of the particle. By way of example we refer to the table below, from a recent work by Palazzi (92), reporting *all* known particles with mass < 1 GeV (excepting gauge quanta and stable leptons) listed in PDG2002:

| Particle | M (MeV) | P | P*u | δ = M − P*u | δ/M (%) | M/P |
|---|---|---|---|---|---|---|
| μ | 105.66 | 3 | 104.38 | 1.28 | 1.21 | 35.219 |
| $\pi^+$ | 139.57 | 4 | 139.17 | 0.40 | 0.29 | 34.893 |
| $\pi^0$ | 134.98 | 4 | 139.17 | −4.19 | −3.10 | 33.744 |
| $K^+$ | 493.68 | 14 | 487.08 | 6.59 | 1.34 | 35.263 |
| $K^0$ | 497.67 | 14 | 487.08 | 10.59 | 2.13 | 35.548 |
| η | 547.30 | 18 | 556.67 | −9.37 | −1.71 | 34.208 |
| ρ | 771.10 | 22 | 765.42 | 5.68 | 0.74 | 35.050 |
| ω | 782.57 | 22 | 765.42 | 17.15 | 2.19 | 35.571 |
| $K^{*+}$ | 891.66 | 26 | 904.58 | −12.92 | −1.45 | 34.295 |
| $K^{*0}$ | 896.10 | 26 | 904.58 | −8.48 | −0.95 | 34.465 |
| p | 938.27 | 27 | 939.38 | −1.10 | −0.12 | 34.751 |
| n | 939.57 | 27 | 939.38 | 0.19 | 0.02 | 34.799 |
| η' | 957.78 | 28 | 974.17 | −16.39 | −1.71 | 34.208 |
| $f_0(980)$ | 980.00 | 28 | 974.17 | 5.83 | 0.60 | 35.000 |
| $a_0(980)$ | 984.70 | 28 | 974.17 | 10.53 | 1.07 | 35.168 |
| Standard deviation | | | | 9.41 | 1.54 | 0.533 |

Table n. 1; From Palazzi [ref. (92), table 1, adapted]

Very careful statistical studies on the $\alpha$-quantization of the masses of elementary particles have been conducted, especially by Mc Gregor and Palazzi, to whose works we refer (92-106). On the other hand, an $\alpha$-quantization of the particle lifetimes also clearly emerges from these works. By way of example, Fig. 2 shows *all* the particles listed by PDG2004 with lifetime > $10^{-21}$ s. In this figure, the ordinate is arbitrary, while the lifetime $\tau$ of each particle is expressed by the abscissa $x_i$, where $\tau = \alpha^{x_i} \tau(\pi^{\pm})$. The lifetimes are clearly, to a very close approximation, expressed by powers of the electromagnetic fine structure constant $\alpha$, regardless of whether we are dealing with leptons, mesons or baryons. We cannot embark on a more detailed discussion here, for which we refer to the original works mentioned above. We shall merely point out that by introducing some "hyperfine structure" corrections, agreement with the power law can be improved considerably.

The existence of the $\alpha$-quantization phenomenon and its universality do not emerge in the conventional theoretical treatment based on the Standard Model. The simple fact that the electromagnetic fine structure constant can govern non electromagnetic decay processes or define the mass of hadrons cannot easily be understood in the context of the conventional framework. The unification of QFT with a theory of relativity such as PSR, within which the microuniverse concept can be clearly formulated, may possibly lead to a deeper understanding of these phenomena.

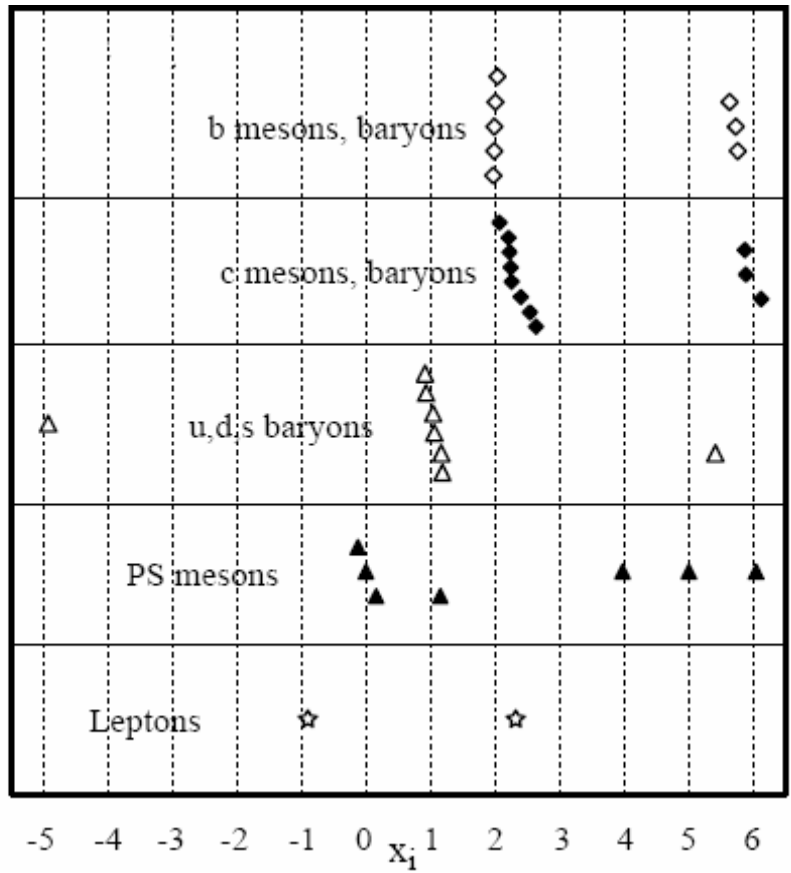

Fig. 2; From Mc Gregor [ref. (100), Fig. 1, adapted]

**4.3 A conjecture on gravitational collapse.**
When a body undergoes a gravitational collapse, the elementary particles of which it is made up are compressed into an extremely small volume. Since, at this time, no physical phenomena are known which can stop the collapse above a certain total mass limit (approximately 2 solar masses for low implosion speeds), this volume will practically be nil. The result is the formation of a singularity of the gravitational tensor $g_{ik}$ (commonly interpreted as "spacetime metric tensor").
However, according to the present approach, leptons, though spatially pointlike, have a time extension of $2\theta \leq 2\theta_0$; as regards hadrons, they are not pointlike, as they have a finite extension $\leq 2c\theta_0$.
It is possible to hypothesize that in both cases gravitational interaction, coupled with the quadrimpulse, is actually coupled with the finite spatial or temporal extension system as a whole. The mass of individual quarks is a model-dependent concept and therefore, to be precise, one cannot talk of a quadrimpulse for an individual quark even if its quadrivelocity is defined.
If this were the case, the gravitational collapse of a body would stop once its radius had fallen below the value $c\theta_0$; thus the volume and density of a body would remain finite and no singularity of $g_{ik}$ would appear. Inside black holes, therefore, not singularities but hyperdense nuclei would be present, capable of developing in their own neighbourhood a gravitational action so intense that it would generate a horizon of the events.

## 5. QFT AND GRAVITY

The search for a quantum theory of gravitation today is essentially focused on the unification of QFT and GR. In this section, based on simple qualitative topics that can be deduced from the ontology presented in section 3, we argue, instead, that the proper quantum theory of gravitation[4] ought to consist of a suitable QFT within the context of PSR.

We point out that, in accordance with this ontology, physical reality is constituted solely by **R** processes, which are the extreme events of transactions. The causal connection between these extremes is not ensured by the propagation of an object in spacetime, be it a wave, a corpuscle or any other entity which acts as an intermediary; rather, it is ensured by the common emergence of the extremes from the same aspatial and atemporal underlying reality. By using a Bohmian terminology (107), we can define this reality as the *background* from whose differentiation the *foreground* constituted by the events emerges.

By attributing the status of physical reality only to **R** processes it follows that the spacetime coordinates are labels associated with these processes and which express certain properties of relation between them. Spacetime as such is only the domain set of these labels and has no physical reality of its own. It is, so to speak, "materialized" by **R** processes.

This approach entails various consequences in the interpretation of QFT formalism and in the search for its possible relations with gravitation.

1) Since no physical reality is attributed neither to **U** processes nor to the empty space in which they are propagated, also the vacuum polarization around a free particle is acknowledged as unreal. This is consistent with the fact that an isolated particle, as it has well defined mass and energy, cannot emit or reabsorb "virtual" particles. The exchange of virtual particles is a feature of interaction processes (70).

2) It follows from the denial of the indipendent physical reality of spacetime that spacetime cannot be a physical agent. Thus either the identification of $g_{ik}$ with the spacetime metric tensor is incorrect or the possibility of describing gravitation by means of this tensor is limited to a non quantum context.

3) The quantization of gravity is normally addressed by assuming that the gravitational field coincides with spacetime metrics. To the possibility of different configurations of the field compatible with the assigned constraints, there corresponds therefore a (quantum) superposition of "metric states". Let us thus consider the case of a particle that interacts solely with the gravitational field, and is therefore "free" in the GR sense. If it interacts with a superposition of metric states, its world line will, in turn, be in a (quantum) superposition of geodesics. In other words, this particle will be dissociated, though free; we are back, therefore, to the incompatibility expressed under point 1).

It seems therefore that the agreement with the ontology expressed in section 3 requires that gravitation be described by means of a QFT against a dynamically inert spacetime background. The $g_{ik}$ tensor ought to be understood in the Cartan sense as a description of the deformation undergone by free motion as a result of the gravitational field, rather than as a spacetime metric tensor. As far as the identification of the geometry of the "background" is concerned, it appears to be strongly constrained by the following obvious requirements:

---

[4] We shall not enter into the matter of its possible structure and refer to ref. (42) for a more in-depth discussion.

a) the background must locally be a Minkowski chronotope and must be four-dimensional;

b) it must admit a global group of transformations in itself and therefore a holonomous geometry;

for only at these conditions can Einstein's principle of relativity be satisfied.
As shown by Fantappié in 1954 (1), the only possibility is thus constituted by a de Sitter chronotope or by its geodesic projection, the Castelnuovo chronotope. The latter is the PSR chronotope.

## 6. CONCLUSIONS

Besides the search for a covariant formalism, the unification of quantum theories with relativity requires the adoption of appropriate ontologies. The transactional ontology seen from a Bohmian viewpoint (summarized in section 3) leads to a new type of nonlocality, which we can define as *cosmological*. This must not be confused with the well known quantum nonlocality, expressed for example in the EPR or GHZ experiments.
On of the aspects of this cosmological nonlocality is the possibility to describe elementary particles as de Sitter-Castelnuovo microuniverses, in a scale ratio with the macrouniverse. This latter is described by the same geometry, because the gravitation can not be a manifestation of spacetime geometry.
In determining this scale ratio a crucial role is played by the electromagnetic fine structure constant, which becomes an important quantization factor within particle microuniverses. This "third quantization", the existence of which is well proved by statistical studies on mass and lifetime spectra of elementary particles, is completely ignored by current QFT, though not in contrast with it.
Acknowledgement of the existence of this additional quantization level, consistent with Heisenberg's insight cited at the beginning of this paper, could be the key not only to solve the problem of divergences in QFT but also that of singularities in gravitational theory.
In any case, the relation between QFT and de Sitter-Fantappié-Arcidiacono relativity deserves careful consideration and appropriate investigation.

**Appendix : The *N* index**

Let us consider an elementary particle of lifetime *T*. Let us suppose this process emerge to physical reality when *N*+1 identical portions of it are concatenated and causally connected. Let $\theta$ be the duration of each portion; the time direction will be the same for all the portions.
If no energy is released to the vacuum in order to generate this causal concatenation, the particle remains at the virtual state. Each portion will manifest itself with a time direction which is the same of the previous portion (probability ½) or the opposite (probability ½). The particle will appears –as a vacuum fluctuation- with a probability $2^{-N}$, and the mean duration of each portion will be :

$$\theta = T \times 2^{-N}.$$

The succession of *N*+1 portions considered as a whole has a duration of $(N+1)\theta$. Indicating with $Mc^2$ the energy needed to induce the time-ordered concatenation, from the uncertainty principle the following relation is derived :

$$(N+1)\theta = \hbar/Mc^2.$$

Therefore :

$$(N+1)/2^N = \hbar/(Mc^2 T), \qquad (A.1)$$

A relation which is very similar to that proposed by Ramanna and Sreekantan (59). The index *N* is not necessarily an integer number and its meaning is rather obvious; in fact :

$$N = \log_2(T/\theta) . \qquad (A.2)$$

In other words, *N* is the information (in bits) associated to the casual appearance of the particle as a vacuum virtual fluctuation.

In Table A1 all the particles with lifetime $T \geq 10^{-21}$ seconds are grouped according to the McGregor scheme (Fig. 2), and the value of *N* (taken as the mean on each single group) is reported in the right column. These mean values are reported in Fig A1 as a function of the progressive number related to each group.
With the exception of two jumps following the neutron (too long lifetime due to phase space reasons) and the pion, *N* decreases for steps of 3 :

neutron → 64 → 61 → 58 → 55 → 52 → 49 → 46 → 43 → pion jump

→ 28 → 25 (hole) → 22 → 19 → 16

The origin of the $\Delta N = 3$ approximated rule is unknown. Speculatively, we can suppose each $\theta$-portion be a Zitterbewegung jump at the speed of light. Because jumps are mutually independent, at least three of them are necessary to the particle in order to probe the three-dimensional space, so defining itself.
Finally, a curious recurrence rule appears which concerns electron → neutron, neutron → muon and tau group → pion jumps. We have: $N$(electron)/$N$(neutron) ≈ 137/96.64 = 1.42 ≈ 3/2; $N$(neutron)/$N$(muon) = 96.64/64.32 ≈ 3/2; $N$(tau group)/$N$(pion) = 44.38/28.94 ≈ 3/2.

Table A1

| Mean life $T$ (sec) | Full width $\Gamma = \hbar/T$ (MeV) | Particle | Mass (MeV) | $\Gamma$/Mass | $N$ (eq. A.1) | $N_{mean} \pm$ SD |
|---|---|---|---|---|---|---|
| 1  8.8570E+02 | 7.4314E-25 | neutron | 939.56 | 7.909E-28 | 96.64 | |
| 2  2.1970E-06 | 2.9960E-16 | muon | 105.66 | 2.835E-18 | 64.32 | |
| 3  5.1800E-08 | 1.2706E-14 | $K_{oL}$ | 497.67 | 2.553E-17 | 61.07 | |
| 4  2.6033E-08 | 2.5283E-14 | $\pi_\pm$ | 139.57 | 1.811E-16 | 58.18 | 59.4 ± 1.5 |
| 5  1.2384E-08 | 5.3149E-14 | $K_\pm$ | 493.68 | 1.076E-16 | 58.95 | |
| 6  2.9000E-10 | 2.2696E-12 | $\Xi_o$ | 1314.83 | 1.726E-15 | 54.84 | |
| 7  2.6320E-10 | 2.5007E-12 | $\Lambda$ | 1115.68 | 2.241E-15 | 54.46 | 54.25 ± 0.5 |
| 8  1.6390E-10 | 4.0159E-12 | $\Xi_-$ | 1321.31 | 3.039E-15 | 54.00 | |
| 9  1.4790E-10 | 4.4503E-12 | $\Sigma_-$ | 1197.44 | 3.716E-15 | 53.70 | |
| 10 8.9530E-11 | 7.3517E-12 | $K_{oS}$ | 497.67 | 1.477E-14 | 51.67 | |
| 11 8.2100E-11 | 8.0170E-12 | $\Omega_-$ | 1672.45 | 4.793E-15 | 53.34 | 52.6 ± 0.85 |
| 12 8.0180E-11 | 8.2090E-12 | $\Sigma_+$ | 1189.37 | 6.902E-15 | 52.79 | |
| 13 1.6710E-12 | 3.9389E-10 | $B_\pm$ | 5279.0 | 7.461E-14 | 49.26 | |
| 14 1.5360E-12 | 4.2851E-10 | $B_o$ | 5279.4 | 8.117E-14 | 49.14 | 49.16 ± 0.09 |
| 15 1.4610E-12 | 4.5051E-10 | $B_{os}$ | 5369.6 | 8.390E-14 | 49.09 | |
| 16 1.3900E-12 | 4.7352E-10 | $\Xi_b$ | ? | | | |
| 17 1.2290E-12 | 5.3556E-10 | $\Lambda_{ob}$ | 5624 | 9.523E-14 | 48.90 | |
| 18 1.0400E-12 | 6.3288E-10 | $D_\pm$ | 1869.3 | 3.386E-13 | 47.00 | |
| 19 4.9000E-13 | 1.3433E-09 | $D_s$ | 1968.5 | 6.824E-13 | 45.97 | |
| 20 4.6000E-13 | 1.4309E-09 | $B_{\pm c}$ | 6400 | 2.236E-13 | 47.63 | 46.48 ± 0.82 |
| 21 4.4200E-13 | 1.4891E-09 | $\Xi_{+c}$ | 2466.3 | 6.005E-13 | 46.16 | |
| 22 4.1030E-13 | 1.6042E-09 | $D_o$ | 1864.5 | 8.604E-13 | 45.62 | |
| 23 2.9060E-13 | 2.2650E-09 | tau | 1777.0 | 1.275E-12 | 45.04 | |
| 24 2.0000E-13 | 3.2910E-09 | $\Lambda_{+c}$ | 2284.9 | 1.440E-12 | 44.86 | 44.38 ± 0.70 |
| 25 1.1200E-13 | 5.8768E-09 | $\Xi_{oc}$ | 2471.8 | 2.377E-12 | 44.11 | |
| 26 6.9000E-14 | 9.5391E-09 | $\Omega_{oc}$ | 2697.5 | 3.536E-12 | 43.52 | |
| 27 8.4000E-17 | 7.8357E-06 | $\pi_o$ | 134.98 | 5.805E-08 | 28.94 | |
| 28 5.1024E-19 | 1.2900E-03 | $\eta$ | 547.30 | 2.357E-06 | 23.29 | |
| 29 7.4000E-20 | 8.8946E-03 | $\Sigma_o$ | 1192.64 | 7.458E-06 | 21.52 | |
| 30 2.5027E-20 | 2.6300E-02 | $\Upsilon 3s$ | 10355.2 | 2.540E-06 | 23.18 | |
| 31 1.5307E-20 | 4.3000E-02 | $\Upsilon 2s$ | 10023.2 | 4.290E-06 | 22.37 | 22.50 ± 0.62 |
| 32 1.2419E-20 | 5.3000E-02 | $\Upsilon 1s$ | 9460.3 | 5.602E-06 | 21.96 | |
| 33 7.2331E-21 | 9.1000E-02 | $J/\psi 1s$ | 3096.87 | 2.938E-05 | 19.40 | |
| 34 6.8564E-21 | 9.6000E-02 | $D^*_{\pm(2010)}$ | 2010.0 | 4.776E-05 | 18.65 | 18.66 ± 0.74 |
| 35 2.3424E-21 | 2.8100E-01 | $J/\psi 2s$ | 3685.96 | 7.623E-05 | 17.92 | |
| 36 3.2585E-21 | 2.0200E-01 | $\eta'$ | 957.78 | 2.109E-04 | 16.32 | |

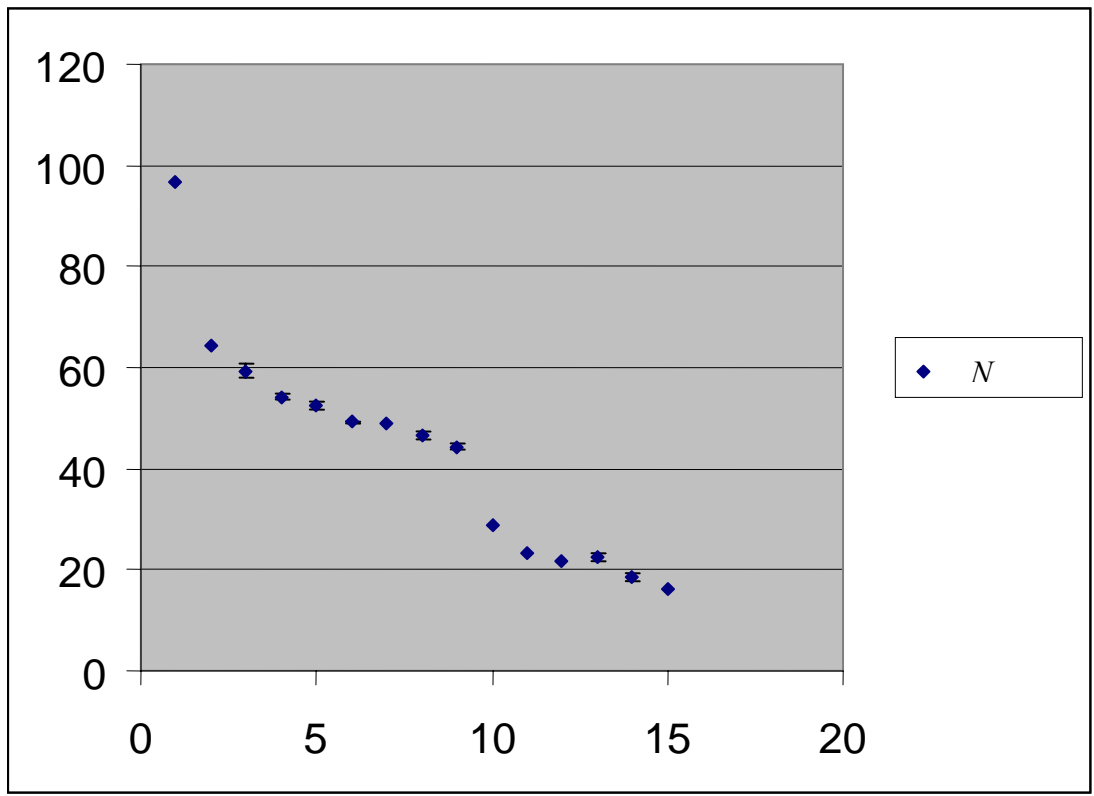

Fig. A1. See text for explanation.

**REFERENCES**

1. Fantappié L.; *Rend. Accad. Lincei* 17, fasc. 5 (1954)
2. Fantappié L.; *Collectanea Mathematica* XI, fasc. 2 (1959)
3. Arcidiacono G.; *Rend. Accad. Lincei* XVIII, fasc. 4 (1955)
4. Arcidiacono G.; *Rend. Accad. Lincei* XVIII, fasc. 5 (1955)
5. Arcidiacono G.; *Rend. Accad. Lincei* XVIII, fasc. 6 (1955)
6. Arcidiacono G.; *Rend. Accad. Lincei* XX, fasc. 4 (1956)
7. Arcidiacono G.; *Rend. Accad. Lincei* XX, fasc. 5 (1956)
8. Arcidiacono G.; *Collectanea Mathematica* X, 85-124 (1958)
9. Arcidiacono G.; *Collectanea Mathematica* XII, 3-23 (1960)
10. Arcidiacono G.; *Collectanea Mathematica* XIX, 51-71 (1968)
11. Arcidiacono G.; *Collectanea Mathematica* XX, 231-255 (1969)
12. Arcidiacono G.; *Collectanea Mathematica* XXIV, 1-25 (1973)
13. Arcidiacono G.; *Collectanea Mathematica* XXV, 159-184 (1974)
14. Arcidiacono G.; *Collectanea Mathematica* XXV, 295-317 (1974)
15. Arcidiacono G.; *Gen. Rel. Grav.* 7, 885-889 (1976)
16. Arcidiacono G.; *Gen. Rel. Grav.* 8, 865-870 (1977)
17. Arcidiacono G.; Projective relativity, Cosmology and Gravitation; Hadronic Press, USA, 1986
18. Arcidiacono G.; The theory of hyperspherical universes; International Center for Comparison and Synthesis, Rome, 1987
19. Arcidiacono G.; La teoria degli universi, vol. II. Di Renzo, Rome 2000 (*in Italian*)
20. Arcidiacono G.; *Collectanea Mathematica* XVI, 149-168 (1964)
21. Arcidiacono G.; *Collectanea Mathematica* XXXIV, 95-107 (1984)

22. Arcidiacono G.; *Boll. Un. Mat.* 2A, 231 (1983)
23. Arcidiacono G.; *Collectanea Mathematica* XXXIV, 197-206 (1984)
24. Arcidiacono G.; *Collectanea Mathematica* XXXIV, 115-129 (1984)
25. Arcidiacono G.; *Hadr. Journ.* 11, 287 (1988)
26. Capelas de Oliveira E., Arcidiacono G., Notte Cuello E.A.; *Hadr. Journ. Suppl.* **13** (3), 249–256 (1998)
27. Gomes D., Capelas de Oliveira E.; *Hadr. Journ. Suppl.* **13** (4), 383–393 (1998)
28. Notte Cuello E.A., Capelas de Oliveira E.; *Int. Journ. Theor. Phys.* **36** (5), 1231–1247 (1997)
29. Notte Cuello E.A., Capelas de Oliveira E.; *Int. Journ. Theor. Phys.* **38** (2), 585–598 (1999)
30. Chernikov N.A., Tagirov E.A.; *Ann. Inst. H. Poincaré Sect. A (N.S.)* **9**, 109–141 (1968)
31. Bros J., Gazeau J.P., Moschella U.; *Phys. Rev. Lett.* **73** (13), 1746–1749 (1994)
32. Takook M.V.; arXiv:gr-qc/0005077 (2000)
33. Hawking S.W., Penrose R.; The Nature of Space and Time; Princeton University Press, 1996
34. Penrose R.; The Emperor's New Mind; Oxford University Press, 1989
35. Penrose R.; The Road to Reality; Oxford, 2004
36. Bohm D.; Wholeness and Implicate Order; Routledge & Kegan Paul, London, 1980
37. Cramer J.G.; *Phys. Rev. D* 22, 362 (1980)
38. Cramer J.G.; *Int. Journ. Theor. Phys.* 27 (2), 227 (1988)
39. Cramer J.G.; *Rev. Mod, Phys.* 58 (3), 647 (1986)
40. Chiatti L.; "Wavefunction structure and transactional interpretation", in "Waves and particles in light and matter", A.van der Merwe and A. Garuccio eds., pp. 181-187. Plenum Press, 1994
41. Chiatti L.; *Found. Phys.* 25 (3), 481 (1995)
42. Chiatti L.; The Archetypal Structures of the Physical World; Di Renzo, Rome, 2005 (*in Italian*)
43. Selleri F., Vigier J.P.; *Nuovo Cim. Lett.* 29, 7 (1980)
44. Bjorken J.D., Drell S.D.; Relativistic Quantum Mechanics; Mc Graw-Hill, New York, 1965
45. Görnitz Th., von Weizsäcker C.F.; *Int. Journ. Theor. Phys.* 27 (2), 237 (1988).
46. Greenberger D.M., Horne M., Zeilinger A.; "Going Beyond Bell's Theorem", in "Bell's Theorem, Quantum Theory and Conceptions of the Universe", M. Kafatos ed., Kluwer, 1989
47. Aspect A. et al. ; *Phys. Rev. Lett.* 49, 91 (1982)
48. Aspect A. et al. ; *Phys. Rev. Lett.* 49, 1804 (1982)
49. Tittel C. et. al.; *Phys. Rev. Lett.* 81, 3563 (1998)
50. Ou Z.Y.; *Phys. Rev. Lett.* 68, 3663 (1992)
51. Schrödinger E.; *Naturwiss.* 49, 53 (1935)
52. Chiatti L.; *EJTP* 15 (4), 17-36 (2007); arXiv:physics/0702178
53. Licata I., Chiatti L.; to be appear in *Int. J. Theor. Phys.*(DOI 10.1007/s10773-008-9874-z, October 2008); arXiv:gr-qc/0808.1339 (2008)
54. Caldirola P.; "Introduzione del cronone nella teoria relativistica dell' elettrone" in "Centenario di Einstein", F. De Finis, M. Pantaleo eds.; Giunti Barbera, Firenze, 1979 (*in Italian*)
55. Sternglass E.J.; *Nuovo Cim. Lett.* 41 (6), 203-208 (1984)
56. Sternglass E.J.; "The Quantum Condition and the non-Euclidean Nature of Space-Time" in : "New Techniques and Ideas in Quantum Measurement Theory"; Annals New York Acad. Sci. 480, 614-617
57. Sternglass E.J.; Before the Big Bang : The Origins of the Universe; Four Walls Eight Windows Publisher, 1998
58. Crosby D.; Space-time Invariance and the Nature of Physics; the work is available on the web
59. Ramanna R., Sreekantan B.V.; *Mod. Phys. Lett.* A10, 741 (1995)
60. Ramanna R.; *Int. Journ. Mod. Phys. A* 11 (28), 5081-5092 (1996)
61. Pati A.K., Jain S.R., Mitra A., Ramanna R.; arXiv:quant-ph/0207144 v1 (2002)
62. Jain S.R., Ramanna R., Ramachandra K.; arXiv:quant-ph/0311063 v1 (2003)

63. Ramanna R., Jain S.R.; *Pramana Journ. of physics* (Indian Academy of Sciences), 57(2-3), 263-269 (2001)
64. Ramanna R., Sharma A.; arXiv:nucl-th/9706063 (1997)
65. Ramanna R., Sharma A.; arXiv:nucl-th/9904068 (1999)
66. Kafatos M.; *Noetic Journal* V.2 (January 1999).
67. Kafatos M.; Bell's Theorem, Quantum Theory and Conceptions of the Universe; Kluwer, Dordrecht, 1989
68. Kafatos M., Nadeau R.; The Conscious Universe: Part and Whole in Modern Physical Theory. Springer Verlag, New York, 1990
69. Kawahara T.; EJTP 15 (4), 37-52 (2007)
70. Chiatti L.; EJTP 10, 33-38 (2006)
71. Caldirola P.; *Nuovo Cim.Suppl.* 3, 297 (1956)
72. Caldirola P. ; *Nuovo Cim. Lett.* 16 (5), 151-155 (1976)
73. Caldirola P., Casati G., Prosperetti A.; *Nuovo Cim.* 43 A, 127 (1978)
74. Caldirola P.; *Nuovo Cim.Lett.* 23, 83 (1978)
75. Caldirola P.; *Nuovo Cim.* 45 A (4), 549-579, (1978)
76. Caldirola P.; *Nuovo Cim.* 49 A (4), 497-511 (1979)
77. Caldirola P.; *Annuario EST Mondadori* , 65-72 (1979)
78. Recami E., Castorina P.; *Nuovo Cim. Lett.* 15 (10), 347 (1976)
79. Caldirola P., Pavsic M., Recami E.; *Nuovo Cim.* 48 B, 205 (1978)
80. Recami E. ; *Annuario EST Mondadori*, 59-64 (1979)
81. Caldirola P., Recami E.; *Nuovo Cim. Lett.* 24 (16), 565 (1979)
82. Recami E.; *Int. Journ. Quant. Chem.* 17, 37-40 (1980)
83. Caldirola P., Maccarrone G.D., Recami E.; *Nuovo Cim. Lett.* 27 (5), 156-160 (1980)
84. Ammiraju P., Recami E.; *Nuovo Cim. Lett.* 78 A (2), 172 (1983)
85. Italiano A., Recami E.; *Nuovo Cim. Lett.* 40 (5), 140 (1984)
86. Recami E., Tonin-Zanchin V.; *Nuovo Saggiatore* 8 , 13 (1992)
87. Recami E., Tonin-Zanchin V.; *Found. Phys. Lett.* 7 (1), 85-93 (1994)
88. Tonin-Zanchin V., Recami E., Roversi J.A., Brasca-Annes L.A.; *Found. Phys. Lett.* 7 (2), 167-179 (1994)
89. Recami E.; "Multi-verses, Micro-universes and Hadrons", preprint NSF-ITP-02-94
90. Hinchliffe I.; Quantum Chromodynamics, chapt. 9; PDG 2002 files, updated to sept. 2001
91. Nambu Y.; *Prog. Theor. Phys.* **7,** 131 (1952).
92. Palazzi P.; http://www.particlez.org/p3a, file 2004-001-v2
93. Palazzi P.; http://www.particlez.org/p3a, file 2004-002-v1
94. Palazzi P.; http://www.particlez.org/p3a, file 2005-001-v2
95. Palazzi P.; http://www.particlez.org/p3a, file 2005-002-v1
96. Palazzi P.; http://www.particlez.org/p3a, file 2005-003-v1
97. Palazzi P.; http://www.particlez.org/p3a, file 2005-005-v1
98. Palazzi P.; http://www.particlez.org/p3a, file 2006-001-v1
99. Palazzi P.; http://www.particlez.org/p3a, file 2006-002-v1
100. McGregor M.H.; *Int. Journ. Mod. Phys.* A 20, 719-798 (2005)
101. Mac Gregor M.H.; *Nuovo Cim. Lett.* **1**, 759 (1971)
102. Mac Gregor M.H.; *Nuovo Cim. Lett.* **4**, 1309-1315 (1970)
103. Mac Gregor M.H.; "Hadron Spectroscopy", in "Fundamental Interactions at High Energy", proceedings of the 1971 Coral Gables Conference; M. Dal Cin, G. J. Iverson eds.
104. Mac Gregor M.H.; *Phys. Rev. D* **9**, 1259-1329 (1974), Sec. XII.
105. Mac Gregor M.H.; *Phys. Rev. D* **13**, 574-590 (1976)
106. Mac Gregor M.H.; *Nuovo Cim.* **103** A, 988-1052 (1990), Sec. 4.
107. Bohm D., Bub J.; *Rev. Mod. Phys.* **38** (3), 453 (1966)